# *Caveats* for the journal and field normalizations in the CWTS ("Leiden") evaluations of research performance

Tobias Opthof [a,b] & Loet Leydesdorff [c]

**Abstract**

The Center for Science and Technology Studies at Leiden University advocates the use of specific normalizations for assessing research performance with reference to a world average. The *Journal Citation Score* (*JCS*) and *Field Citation Score* (*FCS*) are averaged for the research group or individual researcher under study, and then these values are used as denominators of the (mean) *Citations per publication* (*CPP*). Thus, this normalization is based on dividing two averages. This procedure only generates a legitimate indicator in the case of underlying normal distributions. Given the skewed distributions under study, one should average the observed versus expected values which are to be divided first for each publication. We show the effects of the Leiden normalization for a recent evaluation where we happened to have access to the underlying data.

**Keywords**: research performance, evaluation, citation, normalization

[a] Experimental Cardiology Group, Heart Failure Research Center, Academic Medical Center AMC, Meibergdreef 9, 1105 AZ Amsterdam, The Netherlands.
[b] Department of Medical Physiology, University Medical Center Utrecht, Utrecht, The Netherlands.
[c] Amsterdam School of Communications Research (ASCoR), University of Amsterdam, Kloveniersburgwal 48, 1012 CX Amsterdam, The Netherlands; loet@leydesdorff.net; http://www.leydesdorff.net.

# *Caveats* for the journal and field normalizations used in the CWTS ("Leiden") evaluations of research performance

In this study, we take issue with the normalizations proposed in the research evaluations of the Center for Science & Technology Studies (CWTS) at the University of Leiden. These research evaluations are used pervasively and increasingly both in the Netherlands and abroad. In our opinion, the normalizations in these evaluations are seriously flawed because one divides averages instead of averaging divides.

The Center for Science & Technology Studies at Leiden University originated as an offspring of the competition for a department of science dynamics organized by the Dutch government in 1980. The Leiden group specialized in quantitative science studies and joined forces with the Government Council for Science Policy. This new center with specific expertise in scientometric evaluations acquired sufficient funding for licensing the National Citation Reports for the Netherlands as commercially available from the ISI (Thomson-Reuters). Since 1981 an industry for the evaluation of research programs has developed in Leiden that has gained ever more international attention.

Moed *et al*. (1985) reported on the results of the early years. The focus at the time was mainly on descriptive statistics. However, one would like to compare the observed values of numbers of citations with expected values. For this purpose, during the period 1985-1995 the Leiden group developed a set of standards against which the observed values can be evaluated. These standards were derived from the observed set of documents by comparing their mean citation rates with the mean citation rates in the journals and fields of science in which these documents were published. In this sense, the standards are fine-tuned to the micro-diversity prevailing at the level of the research groups to be evaluated and are not based on a one-standard-fits-all criterion. Differences among document types (articles, reviews, and letters) and journals are also taken into account.

Moed *et al.* (1995, at p. 399) explained this normalization as follows:

> "… the number of citations received during the period 1985-1991 by a *letter* published by a group in 1985 in journal X is compared to the average number of citations received during the same period (1985-1991) by all letters published in the same journal (X) in the same year (1985). Generally, a group publishes its papers in several journals rather than one. Therefore we calculated a weighted average JCS indicated as *JCSm,* with the weights determined by the number of papers published in each year."

Analogously to the *JCSm*, a mean citation rate at the field level (*FCSm*) can be calculated on the basis of the attribution of the journals to the subject categories of the ISI database. The mean citation rate of the papers (*CPP*) in the set under study is divided by the *JCSm* and *FCSm* in order to obtain normalization. Values above or below one (1.0) would thus indicate the performance of the group in a standardized way, that is, with reference to the average of the database in terms of journals and fields. *CPP/FCSm* is advocated by the Leiden group as their "crown indicator" (e.g., Van Raan, 2006).

The field normalization is more problematic than the journal normalization because the subject categories of the ISI sometimes heavily overlap and are often misguided (Boyack *et al*., 2005; Leydesdorff & Rafols, 2009; Rafols & Leydesdorff, 2009). These classifications are optimized for the purpose of information retrieval and do not have an analytical base (Pudovkin & Garfield, 2002, at p. 1113n). Boyack *et al*. (2005) estimated that the attribution is correct in approximately 50% of cases across the file (Boyack, personal communication, September 14, 2008). Rafols & Leydesdorff (2009, at p. 1830) noted that some of the subject categories in the biomedical sciences cover virtually the same journals (cosine > 0.95). Because of these overlaps, the subject categories, in our opinion, cannot be used for normalization.

Although Moed *et al*. (1995, at p. 399) admitted that "this classification is far from perfect, but it is at present the only classification available to us," these more recent findings cast serious doubt on the validity of the "crown indicator" (*CPP/FCSm*). Given the sensitivity of evaluations to minor error because of the sometimes relatively small numbers (Laloë & Mosseri, 2009), field normalizations should perhaps be pursued using the classifications of the Library of Congress; these classifications are based on literary warrant and therefore regularly updated (Bensman & Leydesdorff, 2009, at p. 1101). More recently, Bornmann *et al*. (2009. at p. 98) noted that one can nowadays also use the Medical Subject Headings (MeSH) of the bibliographic database MEDLINE (US National Library of Medicine) or the relevant sections of the Chemical Abstracts (CA) database. In contrast to the journal classification scheme of the ISI, MeSH index terms and CA sections are assigned on a paper-by-paper basis. Given the hierarchical structure of the MeSH terms, one can fine-tune the reference set, for example, in the case of the evaluation of highly specialized groups or scholars who publish also in highly-cited journals such as *Nature* and *Science* which are classified by the ISI as "multidisciplinary."

Furthermore, discipline-oriented databases such as Chemical Abstracts, MathSciNet (American Mathematical Society), and PsychINFO (American Psychological Association) have recently extended their records with cited references (Bornmann *et al*., 2009, p. 94). When constructing a reference set, however, a journal or set of journals may not always be the most appropriate unit. New developments, for example, may take place within and/or across journals (Griffith *et al*., 1974; Small, 1978). Garfield's (1972) choice for journal evaluation (using the impact factor) does not imply that the journal is also the most adequate unit of analysis for research evaluation (Bensman, 2007, at pp. 147 ff.). Journals and journal sets—representing specialties—are interwoven in more complex ways (Bradford, 1934; Garfield, 1971; Leydesdorff & Bensman, 2006).

Apart from this problem with the field reference sets, both normalizations (for journals and fields) are perhaps even more seriously affected by the choice of the Leiden team to compare citation rates with the *average* citation rates of journals (*JCSm*) and fields (*FCSm*) instead of averaging the normalizations on a paper-by-paper basis with reference to their specific *JCS* and *FCS* values, respectively. In our opinion, this is a violation of the order of operations which prescribes that divisions precede additions. While one needs a reference set for the normalization (e.g., Schubert & Braun, 1986; Vinkler, 1986, 1997), one should not first mix the apples and the oranges and then compare them with otherwise proportioned sets of apples and/or oranges. When papers in journals such as *Nature* and *Science*—subsumed under the ISI subject category "multidisciplinary"—are mixed in different proportions with papers in highly specialized journals, the reference set itself becomes compounded.

Moed *et al.* (1995, at p. 422) provided the following example of the normalization used by the CWTS:

---

Citation, journal citation and field citation scores

| Paper | Nr. citations | JCS | FCS |
|---|---|---|---|
| I | 17 | 16.9 | 23.7 |
| II | 4 | 3.1 | 3.0 |
| III | 6 | 4.8 | 4.1 |
| IV | 8 | 4.8 | 4.1 |

The average number of citations to the group's *oeuvre (CPP)* is calculated in the following manner:
$$CPP = (17+4+6+8)/(1+1+1+1) = 8.6 \text{ }^a$$
The average citation rate of the journals in which the group has published *(JCSm)* is calculated in the following way:
$$JCSm = [(1 \times 16.9)+(1 \times 3.1)+(2 \times 4.8)]/(1+1+2) = 7.4$$
Similarly, the average citation rate of the subfields in which the group has published *(FCSm)* is given by:
$$FCSm = [(1 \times 23.7)+(1 \times 3.0)+(2 \times 4.1)]/(1+1+2) = 8.7.$$

---

**Table 1**: Counting scheme applied to the calculation of journal and subfield citation rates (Source: Moed *et al.*, 1995, at p. 422). ([a] This value should be 8.75; cf. Van Raan, 2006, at p. 502.)

In the CWTS reports, the *CPP* is divided by the *JCSm* and *FCSm*, respectively, in order to obtain normalized values of 1.18 and 1.00 for this set (Table 2). In our opinion, one should first normalize each row using the number of citations (*C*) divided by the expected citation rate for the journal (*JCS*) or field (*FCS*) and only thereafter average *C/JCS* and

*C/FCS* as shown in the two right-most columns of Table 2.

|  | *Citations* | *JCS* | *FCS* | *C/JCS* | *C/FCS* |
|---|---|---|---|---|---|
| I | 17 | 16.9 | 23.7 | 1.01 | 0.72 |
| II | 4 | 3.1 | 3 | 1.29 | 1.33 |
| III | 6 | 4.8 | 4.1 | 1.25 | 1.46 |
| IV | 8 | 4.8 | 4.1 | 1.67 | 1.95 |
| **Means** | **8.75** *(CPP)* | **7.40** *(JCSm)* | **8.73** *(FCSm)* | *1.30* *(± 0.14)* | *1.37* *(± 0.25)* |
| *CWTS normalization* |  | *1.18* | *1.00* |  |  |

**Table 2**: Counting schemes applied to the calculation of journal and subfield normalizations according to the CWTS *versus* Opthof & Leydesdorff (this study).

Note that our eventual results (in the right-most part of Table 2) are considerably higher than those of the *CWTS* normalizations in this case. We do not first average the number of citations, *JCS*, or *FCS* separately and divide thereafter, but normalize over the rows.

The Leiden method of "addition before division" implies that the actual citations provide an implicit ranking of importance or, in other words, one assumes that more highly cited papers should carry more weight in the index. In our opinion, all papers should have an equal weight in an index. Averaging over the aggregate furthermore allows us to test for the significance of the deviation of the test set from the reference set—we use the standard error of the mean in Table 2, taking into account the number of papers in the two sets. In contrast, using the CWTS procedure—by producing only a single averaged deviation without statistical precision—one cannot distinguish between authors or departments with large and small outputs (CWTS, 2008, at p. 7; cf. Glänzel, 1992).

While one may wish to argue whether the order of operations is applicable to the construction of a specific index, our method takes into account that citations are not normally distributed among papers or within journals and fields. Means are not informative in the case of non-normalized data, whereas the distribution of the normalized data can be approximately normal (and therefore be tested for significance). Furthermore, our proposal for normalization follows the rule for the order of operations which is known by the English mnemonic *Please Excuse My Dear Aunt Sally*: (1) evaluate all expressions with exponents, (2) multiply and divide in order from left to right, and (3) add and subtract in order from left to right (Wu, 2007).[1]

Furthermore, our proposal accords with recent suggestions to collapse citation distributions in different fields by normalizing to the *z*-score (Bornmann & Daniel, 2009) or the relative indicator $c_f = c / c_0$, as proposed by Radicchi *et al*. (2008). In the latter formula, $c$ is the citation rate for a single paper and $c_0$ is the average number of citations per article for the discipline (field, subject area) to which the single paper belongs. In the social sciences, *z*-scores are routinely used to make deviations from the population mean comparable, for example, in the case of intelligence tests. The significance test against the mean in an underlying population, however, should not be confused with the normalization to a (selected!) reference set. Given the skewed distributions, the latter normalization can also—and even more appropriately—be performed using non-parametric statistics such as comparing percentile rank scores with those in the reference set (Plomp, 1990; Pudovkin & Garfield, forthcoming; Rousseau, 2005; cf. Leydesdorff, 1990, 1995).

---

[1] In Dutch, the equivalent mnemonic is known as *Meneer Van Dalen Wacht Op Antwoord.*

In our opinion, the effects of the different normalizations in the example of Table 2 are substantial: not only are this group's scores considerably enhanced, but also and more seriously, the order between the journal and field normalizations is reversed in this example. It would be interesting to see how this might work for larger (e.g., institutionally defined) datasets. However, it is not so easy to replicate the Leiden indicators without having access to the dedicated software of the CWTS. The above example is a fictitious one provided by Moed *et al*. (1995, at p. 422) and repeated by Van Raan (2006, at p. 502). Since the Leiden group operates as a business, the raw data are not publicly available. Customers usually receive normalized averages for management and policy purposes without further analytical details. In the case of a recent evaluation by the CWTS (2008) of research at the Amsterdam Medical Center, however, one of us had access to the data provided by the researchers under study and thus we are able to show the effects of the different normalizations on the evaluation. We should add that neither of us was part of the group being evaluated.

**A real-life example**

Our example is the set of publications by a leading scientist (professor) at the Amsterdam Medical Center who works in biomedical engineering. This subfield includes engineering journals with low impact factors, but the research can also be applied in medical fields with significantly higher expected citation rates. Thus, the author sometimes publishes in high-impact journals and at other times in low-impact ones. Because the distribution of citation scores is therefore very different from a normal one, one can expect large effects from averaging first and dividing thereafter as against first dividing and then averaging. Consequently, the operation is sensitive to Aunt Sally's rule.

Like the CWTS (2008, at p. 67)[2] we found 65 articles by this Principal Investigator during the period 1997-2006,[3] but additionally we noted one review article and two editorials. We agree that editorials can be discarded as so-called "non-citable items" in the ISI database.[4] The report of the CWTS (2008) speaks of "publications" and does not specify the document types. However, to exclude one review article (or any other article?) from an evaluation for unclear reasons is problematic. In our opinion, this is not necessary, because citations to different document types can be scored independently.

In order to remain as close as possible to the CWTS report, we used the 65 articles and proceedings papers and normalized against the sum of the citations to articles and proceedings papers in the respective journals. We used the two categories of document types combined because the distinction between articles and proceedings papers was introduced into the database only in October 2008. (We checked also for normalizing with articles only, and the results were almost identical.) We did not correct for self-citations and limited our analysis to the *JCS*, because the *FCS* cannot be computed without reconstructing the dedicated software of the CWTS. The CWTS (2008) does not provide a value for the *JCSm* including self-citations (*JCSm+*).

---

[2] The report of the CWTS is kept confidential by the AMC because of the personalized information. However, we have obtained permission from the scientist involved to use this example.
[3] The ISI introduced the distinction between articles and conference papers as citable issues in October 2008, and thereafter relabeled articles into these two categories from the perspective of hindsight. We merged these two categories in order to reproduce the CWTS results as closely as possible.
[4] In the context of the impact factor, "citable issues" were defined as articles, reviews, notes, and letters. Notes, however, have been dropped as a category by the ISI in 1997, and articles were divided into articles and proceedings papers in October 2008. However, non-citable issues can be cited and contribute when cited to the numerator of the impact factor, but not to the denominator (Moed & Van Leeuwen, 1996). The focus on citable items has become standard practice in scientometric evaluations.

|  | *CWTS* (2008, at p. 67) | $\Sigma$ observed / $\Sigma$ expected | $\Sigma$ (obs./exp.) |
|---|---|---|---|
| *P(ublications)* | 65 | 65 | |
| *C(itations)* | 679 | 698 | |
| *C/P (CPP+sc)* | 10.45 | 10.74 | |
| without self-citations (*CPP–sc*) | 7.14 | | |
| *JCSm (+sc)* | | 15.23 | |
| *JCSm (–sc)* | 12.22 | | |
| **CPP+/JCSm+** | | **0.71** | **0.91** ($\pm$ 0.11) |
| *CPP–/JCSm–* | **0.58** | | |

**Table 3**: Comparison of the evaluation results with different normalizations for the case of a Principal Investigator in Medical Physics at the Amsterdam Medical Centre.

The results of our reconstruction are summarized in Table 3. Our data show a higher citation rate (698) because of the later date of the measurement (October 2009). Citations accumulate and one is no longer able to measure a previous situation using the *WoS* because the ISI adds journals to the database which are sometimes backtracked into previous years. This policy has equally an upward effect on the *JCSm* (both with and without correction for self-citations).

If the normalization is performed as proposed by us, the score is 0.91 ($\pm$ 0.11) and therewith not significantly different from the world average. The calculation of this value is provided in the Appendix. From the perspective of institutional management, there would in this case have been no reason for concern, while the CWTS evaluation value actually raised some questions. According to the standards of the CWTS (2008, at p. 8), however, the researcher under study would show as performing significantly below the world average in his reference group, both with (0.71) or without self-citations (0.58).

Thus, these evaluations can have unintended effects on the research careers of the persons under study. In our opinion, it is questionable whether an indicator which varies more than 28% when tested for a single scientist, given basic assumptions about the normalization, should be used for managerial or policy purposes. These effects may be very different for different researchers or research groups.[5] Furthermore, let us point to the fact that this difference in absolute terms (0.20) equals the difference between the world average (1.00) and the borderline value of underperformance (0.80) according to the CWTS (2008, at p. 7).

**Ranking the evaluated scientists**

One can wonder whether the differences between our approach and the one used by the CWTS makes a difference to the ranking of these scientists among themselves. Since our group consists of 232 scholars with a total of 12,471 publications (mean = 53.8 ± 48.1 publications/person; median = 41) and we have no automated access to the journal and field normalizations,[6] we decided to study this question in terms of the central issues in any ranking, namely: (1) does an equal ranking in the one (that is, CWTS) rating generate a difference using our model, and (2) if so, how does this difference affect the ranking of high- or low-ranked scientists? Can differences in the ranks be considered significant in terms of the underlying distributions?

---

[5] As noted, one can expect the effect to be smaller the more normal the distribution of the citations over the documents under study.

[6] The citation numbers for the set (CWTS, 2008) are $N = 223,425$; $\bar{x} = 963 \pm 1,138$; median = 566.

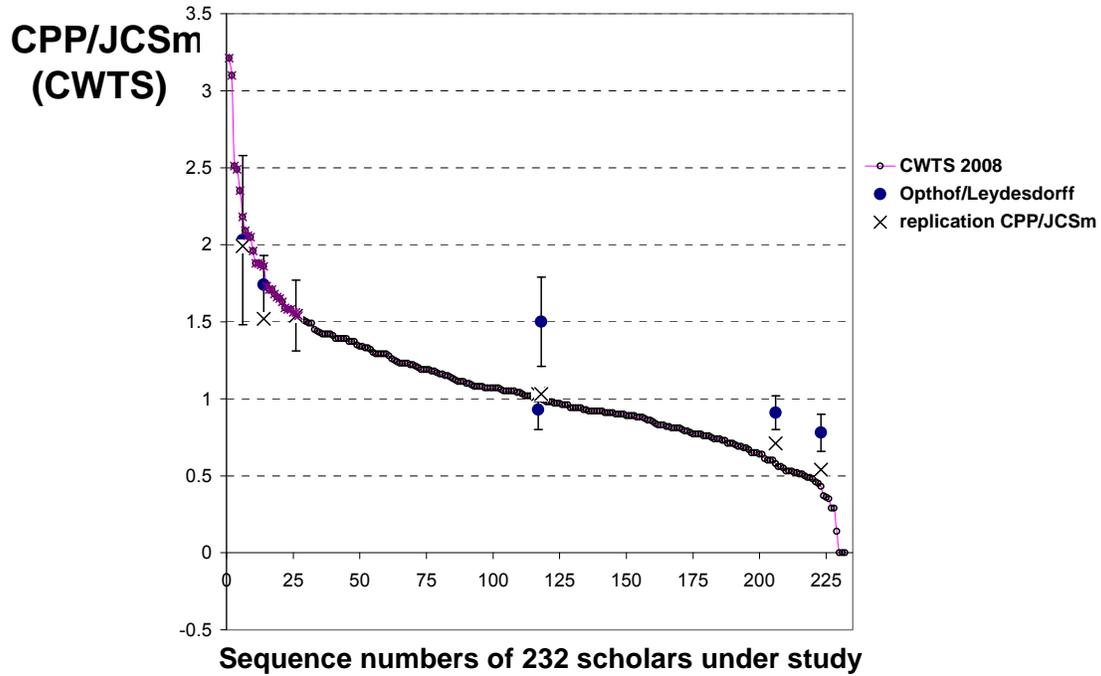

**Figure 1**: Ranking of 232 AMC scientists according to the CWTS (2008); × replication CPP/JCSm in this study; ● average of the observed versus expected values (this study).

| Rank | CPP/JCSm (CWTS, 2008) | N of publications (CWTS, 2008) | N of publications (this study) | Avg(obs/exp) (this study) | CPP/JCSm (replicated) |
|---|---|---|---|---|---|
| 6 | 2.18 | 26 | 23 | 2.03 (± 0.55) | 1.99 |
| 14 | 1.86 | 38 | 37 | 1.74 (± 0.19) | 1.52 |
| 26 | 1.56 | 22 | 22 | 1.54 (± 0.23) | 1.54 |
| 117 | 1.00 | 36 | 32 | 1.50 (± 0.29) | 1.03 |
| 118 | 1.00 | 32 | 37 | 0.93 (± 0.13) | 1.03 |
| 206 | 0.58 | 65 | 65 | 0.91 (± 0.11) | 0.71 |
| 223 | 0.43 | 37 | 32 | 0.78 (± 0.12) | 0.54 |

**Table 4**: The effects of different normalizations on values and ranks.

Figure 1 and Table 4 provide the results of this measurement in seven selected cases. As one would expect, the replication fits the original curve better than our results based on a different indicator ($r = 0.99$ and 0.94, respectively; $p < 0.01$). However there are important differences which our new measure enables us to evaluate quantitatively.

Let us first focus on the 117$^{th}$ and 118$^{th}$ positions, which are tied at *CPP/JCSm* = 1.00 in the CWTS data.[7] While these two authors remain tied at a value of *CPP/JCSm* = 1.03 in the replication of the CWTS results, our results provide 0.93 and 1.50, respectively. Applying Kruskal-Wallis to our data, however, it can be shown that neither of these two distributions of citations is significantly different from 1 (that is, the world average according to the Leiden definition). Additionally, a post-hoc test with Bonferroni correction allows one to compare the mean of the differences between two authors. The difference between the two sets of documents of these authors with respective citation rates is not significant at the 0.05 level. As noted, unlike the CWTS indicators, our results are normal averages and can thus be tested for their significance. Even if one wishes to avoid the assumption of 'normality after normalization', one can still test the differences from 1.00 for each ratio using a non-parametric test.

If we compare the seven cases contained in Table 4 using Kruskal-Wallis, the results are not surprisingly (asymptotically) significant. Testing the distributions for each author against unity reveals that only the second and third authors score significant different from 1, but the first does not. Tested against each other (using the Bonferroni correction), some differences—for example, between the extremes on the ranking—are significant while others are not. The significance of differences depends on the shape of the underlying distributions and the size of the samples, and may not be inferred from inspection of tables and graphs (which do not contain this information).

---

[7] Position 116 is also tied at this value.

For example, the error bars added to Figure 1 would not provide a basis for an inference: the two authors who were tied at 1.00 in the Leiden data obtain very different values using our indicator. Although the error bars do not overlap, this difference is nevertheless not significant. The distribution for one of these two authors is not even significantly different from that of our top author despite the huge differences between them in terms of the Leiden ratings (1.00 and 2.18, respectively).

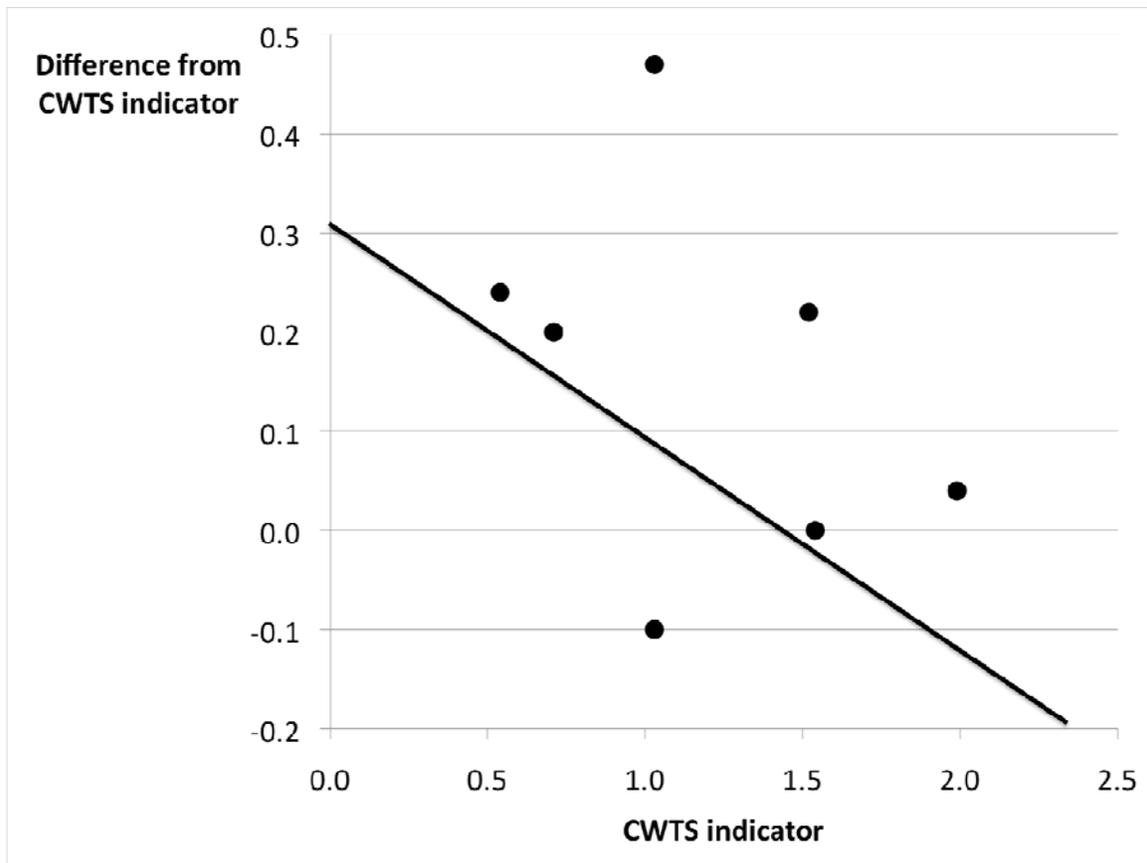

**Figure 2**: The relative difference from the CWTS indicator using our parameter for the seven cases listed in Table 4.

Figure 2 shows that the difference between our method and the CWTS method is not a fixed percentage. The regression line between the CWTS indicator and the deviation from it when using our parameter tends to become larger when the CWTS indicator is lower. Our results suggest a systematic underrating of low-ranked scientists in the CWTS procedure.

In summary, the different normalizations have effects on the rankings in some cases more than others, but neither the differences in rankings nor the effects of the changes are always significant. We understand that research managers and policy makers are focused on rankings in order to legitimize the allocation of resources. However, *caveat emptor*: let the buyer of these evaluations be wary!

**Conclusions and discussion**

We would be the first to agree that any indicator remains a construct that can be used or abused in policy-making and/or managerial processes. However, in our opinion, the quality of an indicator matters nevertheless, and this is particularly urgent in this case because the "Leiden" indicators are increasingly used for the micro-management of research, for example, at the level of faculties and departments. Promotion and funding decisions are increasingly informed by rankings. The Leiden indicators set the standards for the field and are widely used. In summary, the availability of these constructs has an effect on institutional management.

Unfortunately, critical users of the evaluation reports of the CWTS cannot correct for the problems signaled above because of the confidentiality of the information. Users obtain only management information which indicates whether a research group is "above" or "below" average given the applied normalization and summary statistics (Anonymous, 2009). A disaggregated table like the one in Appendix I, however, would allow users to correct for the normalization if they wished. We would like to encourage scientometric evaluators to specify their decisions in using these measurements and normalizations given a tendency among science policy makers and research managers to use these numbers without much further reflection (Bornmann *et al.*, forthcoming). When scientometric indicators are used in the public domain—whether at the macro level of science and technology policy or at the micro level of research management using public (!) funds—the transparency and traceability of these indicators should be one of the primary objectives.

**References**

Anonymus (2009). Advice of the Research Council on the evaluation of the AMC research 2008. Amsterdam: AMC.
Bensman, S. J. (2007). Garfield and the impact factor. *Annual Review of Information Science and Technology,* 41(1), 93-155.
Bensman, S. J., & Leydesdorff, L. (2009). Definition and Identification of Journals as Bibliographic and Subject Entities: Librarianship vs. ISI Journal Citation Reports (JCR) Methods and their Effect on Citation Measures. *Journal of the American Society for Information Science and Technology,* 60(6), 1097-1117.
Bornmann, L., & Daniel, H. D. (2009). Universality of citation distributions—A validation of Radicchi et al.'s relative indicator $c_f = c/c_0$ at the micro level using data from chemistry. *Journal of the American Society for Information Science and Technology,* 60(8), 1664-1670.
Bornmann, L., Leydesdorff, L., & Van den Besselaar, P. (forthcoming). A Meta-evaluation of Scientific Research Proposals: Different Ways of Comparing Rejected to Awarded Applications. *Journal of Informetrics,* in press.

Bornmann, L., Mutz, R., Neuhaus, C., & Daniel, H. D. (2008). Citation counts for research evaluation: standards of good practice for analyzing bibliometric data and presenting and interpreting results. *Ethics in Science and Environmental Politics(ESEP),* 8(1), 93-102.

Boyack, K. W., Klavans, R., & Börner, K. (2005). Mapping the Backbone of Science. *Scientometrics,* 64(3), 351-374.

Bradford, S. C. (1934). Sources of information on specific subjects. *Engineering,* 137, 85-86.

CWTS (2008). AMC-specifieke CWTS-analyse 1997-2006. Leiden: CWTS [access via AMC intranet].

Garfield, E. (1971). The mystery of the transposed journal lists—wherein Bradford's Law of Scattering is generalized according to Garfield's Law of Concentration. *Current Contents,* 3(33), 5–6.

Garfield, E. (1972). Citation Analysis as a Tool in Journal Evaluation. *Science* 178(Number 4060), 471-479.

Glänzel, W. (1992). Publication Dynamics and Citation Impact: A Multi-Dimensional Approach to Scientometric Research Evaluation. In P. Weingart, R. Sehringer & M. Winterhagen (Eds.), *Representations of Science and Technology. Proceedings of the International Conference on Science and Technology Indicators, Bielefeld, 10-12 June 1990* (pp. 209-224). Leiden: DSWO / Leiden University Press.

Griffith, B., Small, H., Stonehil, J., & Dey, S. (1974). Structure of scientific literatures II. Toward a Macrostructure and Microstructure for Science. *Science Studies,* 4(4), 339-365.

Laloë, F., & Mosseri, R. (2009). Bibliometric evaluation of individual researchers: not even right... not even wrong! *Europhysics News,* 40(5), 26-29.

Leydesdorff, L. (1990). Relations Among Science Indicators I. The Static Model,. *Scientometrics 18*, 281-307.

Leydesdorff, L. (1995). *The Challenge of Scientometrics: The development, measurement, and self-organization of scientific communications*. Leiden: DSWO Press, Leiden University; at http://www.universal-publishers.com/book.php?method=ISBN&book=1581126816.

Leydesdorff, L. (2008). *Caveats* for the Use of Citation Indicators in Research and Journal Evaluation. *Journal of the American Society for Information Science and Technology,* 59(2), 278-287.

Leydesdorff, L. (2008). *Caveats* for the Use of Citation Indicators in Research and Journal Evaluation. *Journal of the American Society for Information Science and Technology,* 59(2), 278-287.

Leydesdorff, L., & Bensman, S. J. (2006). Classification and Powerlaws: The logarithmic transformation. *Journal of the American Society for Information Science and Technology,* 57(11), 1470-1486.

Leydesdorff, L., & Rafols, I. (2009). A Global Map of Science Based on the ISI Subject Categories. *Journal of the American Society for Information Science and Technology,* 60(2), 348-362.

Moed, H. F., Burger, W. J. M., Frankfort, J. G., & Van Raan, A. F. J. (1985). The Use of Bibliometric Data for the Measurement of University Research Performance, *Research Policy 14*, 131-149.

Moed, H. F., De Bruin, R. E., & Van Leeuwen, T. N. (1995). New bibliometric tools for the assessment of national research performance: Database description, overview of indicators and first applications. *Scientometrics,* 33(3), 381-422.

Moed, H. F., & Van Leeuwen, T. N. (1996). Impact factors can mislead. *Nature,* 381(6579), 186.

Plomp, R. (1990). The significance of the number of highly cited papers as an indicator of scientific prolificacy. *Scientometrics,* 19(3), 185-197.

Pudovkin, A. I., & Garfield, E. (2002). Algorithmic procedure for finding semantically related journals. *Journal of the American Society for Information Science and Technology,* 53(13), 1113-1119.

Pudovkin, A. I., & Garfield, E. (forthcoming). Percentile Rank and Author Superiority Indexes for Evaluating Individual Journal Articles and the Author's Overall Citation Performance. *CollNet Journal* (In print).

Radicchi, F., Fortunato, S., & Castellano, C. (2008). Universality of citation distributions: Toward an objective measure of scientific impact. *Proceedings of the National Academy of Sciences,* 105(45), 17268.

Rafols, I., & Leydesdorff, L. (2009). Content-based and Algorithmic Classifications of Journals: Perspectives on the Dynamics of Scientific Communication and Indexer Effects *Journal of the American Society for Information Science and Technology,* 60(9), 1823-1835.

Rousseau, R. (2005). Median and percentile impact factors: A set of new indicators. *Scientometrics,* 63(3), 431-441.

Schubert, A., & Braun, T. (1986). Relative indicators and relational charts for comparative assessment of publication output and citation impact. *Scientometrics,* 9(5), 281-291.

Small, H. (1978). Cited documents as concept symbols. *Social Studies of Science* 7, 113-122.

Van Raan, A. F. J. (2006). Comparison of the Hirsch-index with standard bibliometric indicators and with peer judgment for 147 chemistry research groups, *Scientometrics* 67(3), 491-502.

Vinkler, P. (1986). Evaluation of some methods for the relative assessment of scientific publications. *scientometrics,* 10(3), 157-177.

Vinkler, P. (1997). Relations of relative scientometric impact indicators. The relative publication strategy index. *Scientometrics,* 40(1), 163-169.

Wu, H. (2004). "Order of operations" and other oddities in school mathematics; at http://math.berkeley.edu/~wu/order5.pdf (accessed on October 21, 2009).

# Appendix I

| Seq.nr. of publications | Nr. of citations | JCS | CPP/JCS |
|---:|---:|---:|---:|
| 1 | 55 | 58.31 | 0.94 |
| 2 | 46 | 18.46 | 2.49 |
| 3 | 53 | 62.73 | 0.84 |
| 4 | 39 | 48.99 | 0.80 |
| 5 | 24 | 9.98 | 2.40 |
| 6 | 34 | 9.12 | 3.73 |
| 7 | 25 | 20.91 | 1.20 |
| 8 | 18 | 7.52 | 2.39 |
| 9 | 20 | 23.73 | 0.84 |
| 10 | 1 | 1.51 | 0.66 |
| 11 | 14 | 19.74 | 0.71 |
| 12 | 1 | 0.93 | 1.07 |
| 13 | 24 | 17.34 | 1.38 |
| 14 | 23 | 19.10 | 1.20 |
| 15 | 22 | 23.20 | 0.95 |
| 16 | 18 | 45.61 | 0.39 |
| 17 | 11 | 9.98 | 1.10 |
| 18 | 20 | 74.50 | 0.27 |
| 19 | 3 | 1.53 | 1.96 |
| 20 | 3 | 0.61 | 4.95 |
| 21 | 2 | 0.61 | 3.30 |
| 22 | 17 | 65.48 | 0.26 |
| 23 | 14 | 14.32 | 0.98 |
| 24 | 0 | 1.25 | 0.00 |
| 25 | 6 | 7.69 | 0.78 |
| 26 | 12 | 9.98 | 1.20 |
| 27 | 12 | 24.79 | 0.48 |
| 28 | 16 | 19.10 | 0.84 |
| 29 | 11 | 6.41 | 1.72 |
| 30 | 12 | 19.10 | 0.63 |
| 31 | 1 | 0.50 | 1.98 |
| 32 | 11 | 14.32 | 0.77 |
| 33 | 8 | 17.34 | 0.46 |
| 34 | 10 | 7.66 | 1.31 |
| 35 | 9 | 8.01 | 1.12 |
| 36 | 5 | 3.34 | 1.50 |
| 37 | 9 | 14.32 | 0.63 |
| 38 | 8 | 10.16 | 0.79 |
| 39 | 0 | 0.39 | 0.00 |
| 40 | 1 | 3.34 | 0.30 |
| 41 | 6 | 3.34 | 1.79 |
| 42 | 6 | 13.27 | 0.45 |
| 43 | 1 | 3.77 | 0.27 |
| 44 | 6 | 5.61 | 1.07 |
| 45 | 0 | 0.61 | 0.00 |
| 46 | 0 | 0.13 | 0.00 |

| | | | | |
|---|---|---|---|---|
| *47* | 2 | 3.34 | 0.60 | |
| *48* | 4 | 9.98 | 0.40 | |
| *49* | 5 | 23.20 | 0.22 | |
| *50* | 6 | 11.54 | 0.52 | |
| *51* | 1 | 3.34 | 0.30 | |
| *52* | 4 | 7.19 | 0.56 | |
| *53* | 5 | 10.30 | 0.49 | |
| *54* | 2 | 3.34 | 0.60 | |
| *55* | 6 | 17.08 | 0.35 | |
| *56* | 5 | 7.06 | 0.71 | |
| *57* | 4 | 8.61 | 0.46 | |
| *58* | 2 | 23.20 | 0.09 | |
| *59* | 5 | 23.67 | 0.21 | |
| *60* | 3 | 13.95 | 0.22 | |
| *61* | 2 | 8.01 | 0.25 | |
| *62* | 1 | 7.06 | 0.14 | |
| *63* | 2 | 24.07 | 0.08 | |
| *64* | 1 | 18.72 | 0.05 | |
| *65* | 1 | 17.34 | 0.06 | |
| $\Sigma_1^{65}$ | 698 | 989.68 | **0.91** | $= \Sigma_1^{65}(observed/expected)$ <br> *SEM* = 0.11 |
| *CPP/JCSm (CWTS)* | | 698 / 989.68 = **0.71** | | $= \dfrac{\Sigma_1^{65}(observed)}{\Sigma_1^{65}(\exp ected)}$ |

21